\documentstyle[epsfig,amsmath,amssymb,float]{elsart}

\begin{document}
\begin{frontmatter}
\title
{Temperature dependence of the energy of a vortex
in a two-dimensional Bose gas}
\author[sns]{K. K. Rajagopal},
\author[bilal]{B. Tanatar}, 
\author[sns]{P. Vignolo} and 
\author[sns]{M. P. Tosi}
\address[sns]
{NEST and Classe di Scienze, Scuola Normale Superiore, I-56126 Pisa,
Italy}
\address[bilal]
{Department of Physics, Bilkent University, Bilkent, 06800 Ankara,
Turkey}
\maketitle
\begin{abstract}
We evaluate the thermodynamic critical angular velocity $\Omega_c(T)$  
for creation of a vortex of lowest quantized angular momentum in a 
strictly two-dimensional Bose gas at temperature $T$, 
using a mean-field two-fluid model for the condensate and the thermal cloud. 
Our results show that (i) a Thomas-Fermi description of the condensate 
badly fails in predicting the particle density profiles and the energy 
of the vortex as functions of $T$; and (ii) an extrapolation of a 
simple Thomas-Fermi formula for $\Omega_c(0)$ is nevertheless approximately 
useful up to $T\simeq 0.5 T_c$.
\end{abstract}

\end{frontmatter}
\newpage

\section{Introduction}

Quasi-two-dimensional (2D) Bose-Einstein-condensed gases (BEC) have been 
attracting increasing attention in recent years. Flatter and flatter
 condensates have been produced inside magneto-optical harmonic traps 
by squeezing the anisotropy parameter, measured as the ratio 
between the radial and axial trap frequencies 
(see G\"orlitz et al. \cite{Gorlitz2001a} and references therein). As the 
gas approaches the 2D limit, its collisional properties start to influence 
the boson-boson coupling parameter, which becomes dependent on the gas 
density \cite{Morgan2002a,AlKhawaja2002b}. Understanding the 
behaviour of vortices in this regime is important, since they reflect 
the superfluid nature of the 
condensate \cite{Lundh1998a,Castin1999b,Dalfovo2000a} and are expected to 
play a role in the transition from the superfluid to the 
normal state \cite{Kosterlitz1973a}.

An experimental method for the creation of quantized vortices in a 
trapped BEC has made use of an 
"optical spoon" \cite{Madison2000a,Rosenbusch2002a}, whereby the 
condensate in an elongated trap is set into rotation 
by stirring with laser beams. This experiment on a confined boson gas 
is conceptually the analogue of the rotating bucket experiment on 
bulk superfluid Helium, with the difference that 
the thermal cloud is inhomogeneously distributed and lies mostly 
outside the condensate cloud. 
A quantized vortex first appears in the condensate at a critical 
angular frequency of stirring which corresponds to an instability of 
the vortex-free state. A lower bound for the critical 
frequency can be assessed from the energy of a vortex, defined as the 
difference in the internal energy of the gas with and without a vortex. 
Observation of a vortex in a trapped gas is difficult 
since the vortex core is small in comparison to the size of the boson cloud, 
but the size of the core increases during free expansion and indeed 
vortices were first observed experimentally by releasing the trap and 
allowing ballistic expansion of the cloud \cite{Madison2000a}.

In this Letter we consider a 2D rotating condensate at finite temperature 
and evaluate the particle density profiles and the energy of a vortex 
within a strictly 2D model for the boson-boson coupling. This is 
appropriate to a situation in which the {\it s}-wave scattering 
length starts to 
exceed the vertical confinement length. Previous work has established that 
the dimensionality of the scattering collisions strongly affects the 
equilibrium density profiles \cite{Tanatar2002a,Rajagopal2004a} and the 
process of free expansion of a BEC containing a vortex \cite{Hosten2003a}. 
A similar study of the density profiles of a rotating BEC in 3D 
geometry at finite temperature has been carried out by Mizushima {\it et al.} 
\cite{Mizushima2001a}, who also determined the location of various 
dynamical instabilities within the Bogoliubov-Popov theory.
	
We begin, therefore, by introducing our description of a 2D BEC 
containing a vortex at finite temperature $T$. This uses a mean-field 
two-fluid model for the condensate and the thermal cloud.

\section{The model}
The BEC is subject to an anisotropic harmonic confinement characterized 
by the radial trap frequency $\omega_\perp$  and by the axial frequency 
$\lambda\omega_\perp$  with $\lambda\gg 1$. Motions along the $z$ 
direction are suppressed and the condensate wave function is determined 
by a 2D equation of motion in the $\{x,y\}$ plane.

The order parameter for a 2D condensate accommodating a quantized vortex 
state of angular momentum $\hbar\kappa$ per particle is written as 
$\Phi({\bf r})=\psi(r)\exp(i\kappa\phi)$, with $\phi$ the azimuthal 
angle. The wave function $\psi(r)$ then obeys the nonlinear Schr\"odinger 
equation (NLSE)
\begin{equation}
\left[-{\hbar^2\over 2m}
\nabla^2+{\hbar^2\kappa^2\over 2m r^2}+{1\over 2}m\omega_\perp^2r^2+
g_2n_c(r)+2g_1n_T(r)\right]\psi(r)=\mu\psi(r)\, 
\label{nlse}
\end{equation}
where $\mu$ is the chemical potential, $m$ the atomic mass, 
$n_c(r)=|\psi(r)|^2$  the condensate density, and $n_T(r)$
the density distribution of the thermal cloud. 
The 2D coupling parameters $g_j$, with $j = 2$ 
for condensate-condensate repulsions and $j = 1$ for 
condensate-noncondensate repulsions, are given by
\begin{equation}
g_j=\frac{4\pi\hbar^2/m}{\ln|{4\hbar^2/(jm\mu a^2)}|}\, ,
\label{coupling}
\end{equation}
where $a$ is the {\it s}-wave scattering length \cite{Morgan2002a,AlKhawaja2002b}. In Eq. (\ref{coupling}) we have omitted a term due to thermal 
excitations \cite{Rajagopal2004a}, which is negligible in the temperature 
range of present interest ($T\le 0.5 T_c$, 
with $T_c$ the critical temperature).

In our model the atoms in the thermal cloud are not put directly 
into rotation, but feel the 
rotating condensate through the mean-field interactions. 
In the Hartree-Fock approximation ~\cite{Minguzzi1997a} 
the thermal cloud is treated as an ideal gas subject to the effective potential
\begin{equation}
V_{eff}(r)=\frac{1}{2}m\omega_\perp^2r^2+2g_1n_c(r)\,.
\label{eff-pot}
\end{equation} 
The density distribution of the thermal cloud is then given by
\begin{equation}
n_{T}(r)=-\frac{m}{2 \pi \hbar^2\beta} \ln \left\{1-
\exp\left[\beta\left(\mu - V_{eff}(r)- 
\frac{p_{0}^2}{2m}\right)\right]\right\},
\label{thermal}
\end{equation}
with $\beta=1/(k_BT)$.
The momentum cut-off $p_0$ in Eq.~(\ref{thermal}) is a simple expedient
to eliminate discontinuities in the density profiles that can occur
near the Thomas-Fermi radius of the condensate. We take
$p_0=2\sqrt{mg_{1}n_{T}}$ (see for instance \cite{Prokofev2001a}) which 
is equivalent to adding
the term $2g_1n_T$ to the effective potential in Eq. (\ref{eff-pot}).

We solve self-consistently the coupled Eqs. (\ref{nlse})-(\ref{thermal})
together with the condition that  
the areal integral of $n_{c}(r)$+$n_{T}(r)$ is equal to the
total number $N$ of particles.
The differential 
equation (\ref{nlse}) is solved iteratively by discretization, using a 
two-step Crank-Nicholson scheme \cite{Press1992a}. In Sec.~\ref{result} 
we shall also compare the results with those obtained in the Thomas-Fermi 
approximation by dropping the radial kinetic energy term in Eq. (\ref{nlse}).

\subsection{The energy of a vortex}
The critical angular velocity
measured in experiments where vortex nucleation occurs from a 
dynamical instability \cite{Madison2000a} depends strongly on the shape 
of the perturbation.
However, a lower bound for the angular velocity required to produce 
a single-vortex state can be 
estimated once the energies of the states with and without the vortex 
are known.
Since the angular momentum per 
particle is $\hbar\kappa$, the critical (thermodynamic) angular velocity
is given by 
\begin{equation}
\Omega_c=\frac{E_\kappa-E_{\kappa=0}}{N\hbar\kappa}\,.
\label{critfreq}
\end{equation}
This expression follows by equating the energy of the vortex state in 
the rotating frame, that is $E_\kappa-\Omega_cL_z$, to the 
energy $E_{\kappa=0}$ of the vortex-free state.

In the noninteracting case at zero temperature, 
the energy difference per particle is simply 
$\hbar\kappa\omega_\perp$, so that $\Omega_c$ is just the 
trap frequency in the $\{x,y\}$ plane.
For the interacting gas at zero 
temperature in the Thomas-Fermi approximation, Eq. (\ref{critfreq}) 
reduces to the expression \cite{Baym1996a,Lundh1998a}
\begin{equation}
\Omega_c^{TF}(0)=\frac{2\hbar}{m R^2}\ln\left(\frac{0.888\,R}{\xi}\right)\,.
\label{formula}
\end{equation}
Here $R=(2\mu/m\omega_\perp^2)^{1/2}$ and
$\xi=R \,\hbar\omega_\perp/2\mu$ are the Thomas-Fermi radius and the healing 
length respectively.

In the general case of an interacting gas at finite temperature, we have 
to evaluate numerically the total energy as the sum of four 
terms~\cite{Giorgini1997b},
\begin{equation}
E_\kappa=E_{\rm kin,c}+E_{\rm trap}+E_{\rm int}+E_{\rm kin,T}\,.
\end{equation}
These terms are the kinetic energy of the condensate
\begin{equation}
E_{\rm kin,c}=\int d^2r \,\psi^*(r)\left(-{\hbar^2\over 2m}
\nabla^2+{\hbar^2\kappa^2\over 2m r^2}\right)\psi(r)\,,
\end{equation}
the energy of confinement
\begin{equation}
E_{\rm trap}=\frac{1}{2}m\omega_\perp^2
\int d^2r\,r^2[n_c(r)+n_T(r)]\,,
\end{equation}
the interaction energy
\begin{equation}
E_{\rm int}=\frac{1}{2}\int d^2r[g_2n^2_c(r)+4g_1n_c(r)n_T(r)+2g_1n_T^2(r)]\,,
\end{equation}
and the kinetic energy of the thermal cloud
\begin{equation}
E_{\rm kin,T}=\int d^2r \int \frac{dp}{2\pi\hbar^2} \frac{p^3}{2m}
\left\{\exp[\beta(p^2/2m+V_{eff}(r)-\mu)]-1\right\}^{-1}.
\end{equation}
In Sec.~\ref{result} we compare the angular frequency obtained by the 
calculation of the total energy of the 
gas with and without a vortex with those obtained from a Thomas-Fermi 
calculation and from 
extrapolating Eq. (\ref{formula}) at finite temperature through 
the temperature dependence of the chemical potential.

\section{Results and discussion}
\label{result}
For a numerical illustration we have taken $\kappa = 1$ and chosen values 
of the system parameters as 
appropriate to the gas of $^{23}$Na atoms studied in the experiments of 
G\"orlitz {\it et al.} \cite{Gorlitz2001a}, namely    
$\omega_\perp= 2\times188.4$ Hz, $a = 2.8$ nm, and $N = 5\times10^3$. 
We are implicitly assuming, however, that the trap 
has been axially squeezed to reach the strictly 2D scattering regime.

\begin{figure}[h]
\centering{
\epsfig{file=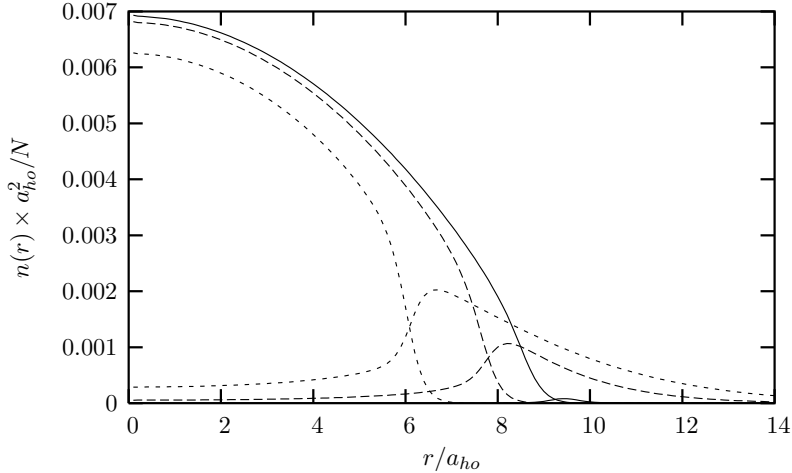,width=0.8\linewidth}}
\caption{Density profiles $n(r)$ for the condensate and the thermal cloud 
(in units of $a_{ho}^2/N$, with $a_{ho}=\sqrt{\hbar/m\omega_\perp}$) 
{\it versus} radial distance $r$ (in units of $a_{ho}$) 
at temperature $T/T_c= 0.05$ (full line), 0.25 (long-dashed line), 
and 0.50 (short-dashed line).}
\label{fig1}
\end{figure}
\begin{figure}[h]
\centering{
\epsfig{file=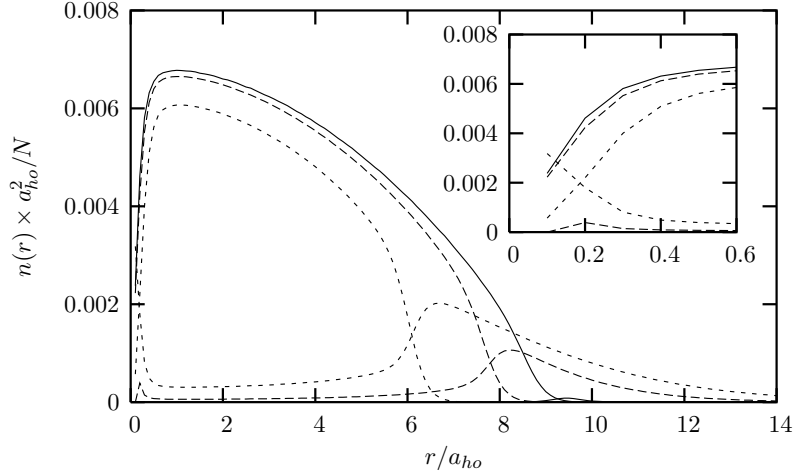,width=0.8\linewidth}}
\caption{Density profiles for the condensate and for the thermal cloud 
in the presence of a vortex. Units and symbols are as in Fig. \ref{fig1}.
The inset shows an enlarged view of the 
profiles near the center of the trap.}
\label{fig2}
\end{figure}
The density profiles obtained from Eqs. (\ref{nlse})-(\ref{thermal})
for the gas at three different values of the temperature  
(in units of the critical temperature 
$T_c=(\sqrt{6N}/\pi k_B)\hbar\omega_\perp$ of the ideal Bose gas) 
are shown in Figs. \ref{fig1} and \ref{fig2}. In the absence of a vortex 
(Fig. \ref{fig1}), the main point to notice is that the 
growth of the thermal cloud exerts an increasing repulsion on the 
outer parts of the condensate, constricting it towards the central 
region of the trap. This effect is seen only when the condensate 
is treated by the NLSE and persists in the presence of a vortex 
(Fig. \ref{fig2}) but , as we 
shall see below, is missed in the Thomas-Fermi approximation where 
the radial kinetic energy of 
the condensate is neglected. In addition, the thermal cloud penetrates 
the core of the vortex and enhances the expulsion of the condensate 
from the core region (inset of Fig. \ref{fig2}), in a manner which 
again is governed in its details by the radial kinetic energy term in the NLSE.

\begin{figure}[h]
\centering{\epsfig{file=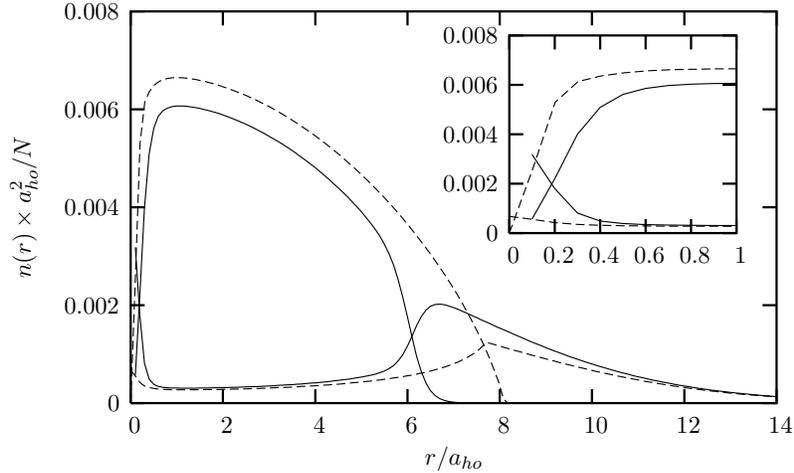,width=0.8\linewidth}}
\caption{Density profiles for a BEC containing a vortex at $T/T_c= 0.50$ 
in the full calculation using the NLSE (full lines) and in the 
Thomas-Fermi approximation (dashed lines). The units 
are as in Fig. \ref{fig1}. The inset shows an enlarged view of the 
profiles near the center of the trap.}
\label{fig3}
\end{figure}
The profiles obtained for a BEC containing a vortex at $T = 0.5\,T_c$  
by the full numerical calculation using the NLSE are compared in 
Fig. \ref{fig3} with those obtained in the Thomas-Fermi 
approximation. The constriction of the condensate in its outer parts 
and its expulsion from the core region by the thermal cloud are clearly 
underestimated in the Thomas-Fermi theory.

\begin{figure}[h]
\centering{\epsfig{file=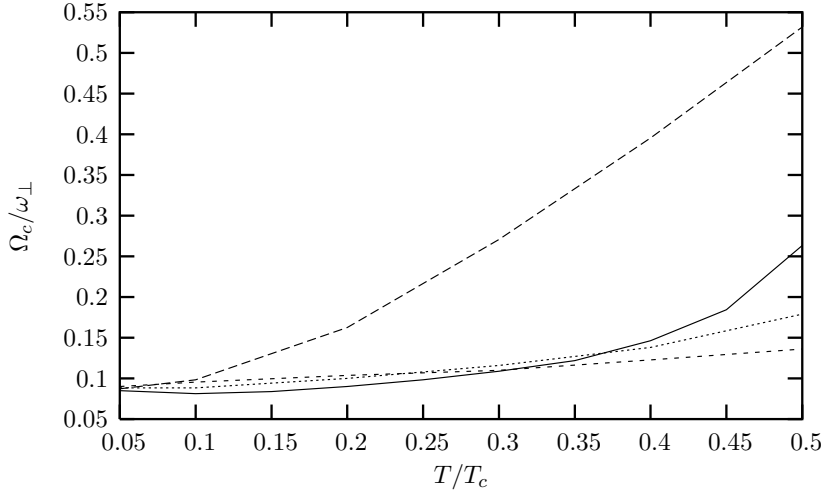,width=0.8\linewidth}}
\caption{The energy of a vortex, expressed as the thermodynamic critical 
frequency $\Omega_c$ in units of 
the radial trap frequency $\omega_\perp$, as a function of $T/T_c$. 
The results from the full calculation using the NLSE (full line) are 
compared with those obtained from Eq. (\ref{formula}) with the corresponding 
values of $\mu(T)$ (dotted line) and with the Thomas-Fermi values of
$\mu(T)$ (short-dashed line). 
The long-dashed line shows the results obtained in a calculation 
using the Thomas-Fermi theory.}
\label{fig4}
\end{figure}	
Finally, Fig. \ref{fig4} reports our results for the energy of the 
vortex as a function of $T/T_c$. The 
inaccuracies arising in the density profiles from the Thomas-Fermi 
treatment of the interplay between the condensate and the thermal 
cloud clearly lead to large errors in the estimation of $\Omega_c(T)$.

On the other hand the simple expression given in Eq. (\ref{formula}), 
with the two alternatives of using in it the chemical potential $\mu(T)$  
from the Thomas-Fermi approximation or from the full 
calculation using the NLSE, gives a reasonable account of the vortex 
energy up to $T\simeq0.5\,T_c$.

\section{Summary and future directions}

In summary, we have calculated the density profiles and the thermodynamic 
critical frequency for vortex nucleation in a strictly 2D 
Bose-Einstein-condensed gas at various temperatures. Our 
calculations have demonstrated the interplay between the thermal cloud 
and the structure of the condensate in the regions where the radial 
kinetic energy of the condensate is playing an important role 
(namely, in the outer parts of the condensate and in the vortex core) 
and validated the use of a simple expression of the vortex energy for 
temperatures up to about 0.5 $T_c$.

As to future directions, a practical realization of the 2D regime that 
we have investigated would require a combination of axial squeezing 
of the trap and an enhancement of the scattering length. 
In relation to the latter, a treatment of a Feshbach resonance in a 2D gas 
will be reported shortly.

\ack
This work has been partially supported by an Advanced Research Initiative 
of SNS. B.T. acknowledges support from TUBITAK, TUBA, and thanks SNS 
for hospitality and INFM for financial support 
during part of this work.

%\bibliographystyle{elsart-num.bst}
%\bibliography{pr}

\end{document}